\documentclass[prd,12pt,tightenlines,nofootinbib,groupedaddress]{revtex4}
\usepackage{amsmath,amsfonts,amssymb,amsthm,bbm,psfrag}
\usepackage{subfigure}
\usepackage[colorlinks,citecolor=blue,linkcolor=blue,urlcolor=blue]{hyperref}
\usepackage{mathrsfs}
\usepackage{titlesec} 
\usepackage[T1]{fontenc}
\usepackage{latexsym}
\usepackage{amsbsy}
\usepackage{mathrsfs}
\usepackage{xcolor}
\usepackage{enumerate}
\usepackage{calc}
\usepackage{latexsym}
\usepackage{graphicx,calc,epsfig}
\usepackage{mathtools}

\newcommand{\Ref}[1]{(\ref{#1})}

\newcommand{\scr}{\scriptscriptstyle\rm}

\newcommand{\cL}{{\mathcal L}}

\newcommand{\be}{\begin{equation}}
\newcommand{\ee}{\end{equation}}
\newcommand{\bea}{\begin{eqnarray}}
\newcommand{\eea}{\end{eqnarray}}
\newcommand{\bs}{\begin{subequations}}
\newcommand{\es}{\end{subequations}}
\newcommand{\nn}{\nonumber}

\newcommand{\w}{\wedge}

\newcommand{\f}{\frac}
\newcommand{\tl}{\tilde}
\def\p{\partial}

\newcommand{\na}{\nabla}

\def\a{\alpha}
\def\b{\beta}
\def\g{\gamma}
\def\d{\delta}

\def\eps{\epsilon}

\def\th{\theta}

\def\k{\kappa}
\def\l{\lambda}
\def\m{\mu}
\def\n{\nu}

\def\r{\rho}

\def\s{\sigma}
\def\t{\tau}

\def\om{\omega}
\def\G{\Gamma}
\def\D{\Delta}

\def\L{\Lambda}

\newcommand{\og}[1]{\overset{\scriptscriptstyle e}{#1}{}}
\newcommand{\ut}[1]{ \underset{\widetilde{}}{#1}{} }

\renewcommand\thesection{\Roman{section}}
\titleformat{\section}{\large\scshape\bfseries\centering}{\thesection.}{.7em}{}
\titleformat{\subsection}{\scshape\bfseries}{\thesubsection.}{.7em}{}

\begin{document}

\title{\large\scshape\bfseries Spacetime Thermodynamics with Contorsion}

\author{Tommaso De Lorenzo$^1$, Elena De Paoli$^{1,2}$ and Simone Speziale$^1$}
\affiliation{$^1$ Aix Marseille Univ., Univ. de Toulon, CNRS, CPT, UMR 7332, 13288 Marseille, France \\
$^2$ Dip. di Fisica, Univ. di Roma 3, Via della Vasca Navale 84, 00146 Roma, Italy} 

\begin{abstract}

We prove that a conserved effective energy-momentum tensor for Einstein-Cartan theory can be identified from the Noether identities of the matter Lagrangian, using the torsion field equations relating them. 
More precisely, a one-parameter family labelled by the Immirzi parameter.
We use this result and the contorsion description to show that Jacobson's thermodynamical derivation of the Einstein equations follows as in the metric theory, namely from the equilibrium Clausius relation and the fact that a Killing horizon is metric-geodetic. Our derivation works for an arbitrary torsion field.
In the course of our discussion we review the laws of black hole mechanics and their dependence on torsion.
\end{abstract}

\date{\today}
\maketitle
\tableofcontents

\section{ Introduction}

In a famous paper \cite{Jacobson:1995ab}, Ted Jacobson proposed that Einstein equations could have a thermodynamical origin, compatible with the thermodynamical interpretation of the laws of black hole mechanics \cite{Bardeen:1973gs}. His argument, based on a geometric interpretation of Clausius relation,
has been later extended to include non-equilibrium terms and higher derivative gravity theories \cite{Eling:2006aw,Chirco:2009dc,Guedens:2011dy}, and more recently to spacetimes with non-propagating torsion, namely Einstein-Cartan first-order gravity \cite{Dey:2017fld}.
This last paper motivates the study presented here.
The main difficulty of extending Jacobson's idea to Einstein-Cartan gravity is that there are two sets of independent field equations to be derived: the torsion equations as well as the Einstein equations proper.
The authors of \cite{Dey:2017fld} show that it is possible to derive the latter set for a special type of torsion, and by identifying the torsional terms as a non-equilibrium contribution to Clausius relation. The torsion equations are not derived, and whether they can have also a thermodynamical origin is left as an open question. In our paper we also do not provide a derivation of the torsional equations, but we show that if they hold, the tetrad Einstein equations can be derived without the need of non-equilibrium terms nor restrictions on torsion. The technical result that allows us to achieve this is the identification of the conserved energy-momentum tensor.

The last point is crucial: in Einstein-Cartan theory, there is no conserved energy-momentum tensor that appears as source of the field equations. Nonetheless, if one restricts to invertible tetrads (and this appears necessary to connect with the metric theory and the familiar notions used in Jacobson's argument), the connection can always be written as a Levi-Civita one plus a contorsion tensor. Using this well-known decomposition, the tetrad Einstein equations can be written as the Levi-Civita Einstein tensor on the left hand side, and a torsion dependent effective energy-momentum tensor $T^{\rm eff}$ on the right hand side. By taking the Levi-Civita covariant derivate of both sides, the left one vanishes due to Bianchi's identities. This in turn implies the vanishing of the right hand side, allowing to identify the conserved energy-momentum tensor also in the presence of torsion. 
For the thermodynamical argument, on the other hand, one needs to identify a conserved energy-momentum tensor without using the field equations, since these are to be derived. The first result of our paper is to show that the conservation of $T^{\rm eff}$ in the Einstein-Cartan theory can be derived \emph{without} using the tetrad field equations. The proof is simple although rather lengthy, and best done using differential forms. It follows from the Noether identities of the theory, and requires the matter and torsion field equations to be satisfied.

Our second result is to use this conserved energy-momentum tensor and the contorsion description to show that the tetrad Einstein equations can be derived from the Clausius relation with the same assumptions and hypothesis of the metric case \cite{Jacobson:1995ab}, without the need of the non-equilibrium terms and the  restrictions on torsion used in \cite{Dey:2017fld}. 
This is possible because the starting point of Jacobson's argument, a Killing horizon associated with a locally boosted observer, is a notion which is insensitive to the presence of torsion. 
In particular, the generators of the Killing horizons follow the Levi-Civita geodesic equation. This turns out to suffice to recover the tetrad Einstein equations from the equilibrium Clausius relation, since the torsion terms are identified by the effective energy-momentum tensor.
A further advantage of our derivation is that it includes also the Immirzi term in the Einstein-Cartan theory.
Our results build on the discussion of \cite{Dey:2017fld}, albeit with a critique of some of their methods and results. 
Our title is motivated by this paper, and meant to stress the role that the contorsion decomposition plays in the derivation.

To complete our discussion, we also look at the laws of black hole mechanics in the presence of torsion. The zeroth law is unaffected, and it can be proven exactly as in the metric case, provided that the energy conditions are imposed on $T^{\rm eff}$. The first law on the other hand depends on torsion. We consider here the `physical process' version of the first law \cite{Wald:1995yp}, which is closely related to Jacobson's argument run backwards. Using the same contorsion decomposition as before, the formal expression of the first law is unchanged, but the quantities appearing depend on torsion through the effective energy-momentum tensor.
The second law has a more marginal dependence, in the sense that torsion simply enters the inequalities on the energy conditions required.

Finally we give in Appendix \Ref{AppTeodoro} a brief comparison of two slightly different versions of Jacobson's argument \cite{Jacobson:1995ab,Guedens:2011dy}, and present an alternative derivation technically closer to the first law.

We use metric signature with mostly plus, and natural units $G=c=\hbar=1$.

\section{Einstein-Cartan field equations and matter sources}
Let us begin by briefly reviewing the field equations of Einstein-Cartan theory and the contorsion decomposition. We refer the reader to \cite{Hehl:1994ue} for more details, and to the Appendix~\ref{AppA} for definitions and notation.
We consider the following first-order action, 
\be\label{SEC}
S_{\scr EC}(e,\om) = \f1{16\pi\,} \int P_{IJKL} \Big(e^I\w e^J\w F^{KL}(\om) - \frac\L6 e^I\w e^J\w e^K\w e^L\Big),
\ee
where
\be
P_{IJKL} := \f1{2\g}(\eta_{IK}\eta_{JL}-\eta_{IL}\eta_{JK}) +\f12\eps_{IJKL},
\ee
and $\g$ is the Immirzi parameter. 
We restrict attention to invertible, right-handed tetrads. The action is then equivalent to first-order general relativity \footnote{Sometimes called Einstein-Palatini general relativity because proving its equivalence to general relativity uses the Palatini identity.} 
\be
S_{\scr EP}(g,\G) = \f1{16\pi\,} \int [\sqrt{-g}(g^{\m\r}g^{\n\s}R_{\m\n\r\s}(\G) - 2\L) +\f1\g\tl\eps^{\m\n\r\s}R_{\m\n\r\s}]d^4x,
\ee
with initially independent metric and connections, which are related to the fields of \Ref{SEC} by  the familiar formulas
\be\label{gG}
g_{\m\n}=e_\m^I e_\n^J\eta_{IJ}, \qquad \G^\r_{\m\n} = e_I^\r D_\m e^I_\n:=e_I^\r (\p_\m e^I_\n+\om^{IJ}_\m e_{J\n}).
\ee

We collectively denote the matter fields as $\psi$, and consider a general matter 
Lagrangian $L_m(e,\om,\psi):={\cal L}_m(e,\om,\psi) d^4x$.
Varying the matter action we have
\begin{align}\label{dSm}
&\d S_m = \int \d L_m= \int \left(2 \tau^\m{}_I \d e^I_\m + \s^\m{}_{IJ} \d\om^{IJ}_\m + E_m\d\psi\right) e \, d^4x,
\end{align}
where $E_m$ denotes the matter field equations, and we defined the source terms
\begin{align}
&\tau^\m{}_I := \f1{2e}\f{\d \cL_m}{\d e^I_\m}= - \f1{2e}\f{\d \cL_m}{\d e^{\n}_J} e^\n_I e^\m_J =: \t^J{}_\n e^\n_I e^\m_J, 
\qquad \s^\m{}_{IJ} = \f1{e}\f{\d \cL_m} {\d \om^{IJ}_{\m}}. \label{deftausigma}
\end{align}
The sign choice in the definition of $\t$ is not universal in the literature. We picked it this way in analogy with the metric energy-momentum tensor $T_{\m\n}^\G$,  
\begin{align}\label{Tt}
&T^{\G}_{\m\n} := -\f2{\sqrt{-g}} \f{\d \cL_m(g,\G)}{\d g^{\m\n}}  = - \f1{e} \f{\d \cL_m(e,\om)}{\d e^{I(\m}}e^I_{\n)} = 2\t^I{}_{(\m} e_{\n)I},
\end{align}
which coincides with the one of general relativity in the absence of torsion.

The field equations obtaining varying \Ref{SEC} and the matter action are 
\begin{subequations}\label{FE}\begin{align}\label{FEE}
& G^\m{}_I(e,\om) +\L e^\m_I +\f1{2\g } \eps^{\m\n\r\s}e^\a_I R_{\a\n\r\s}(e,\om) =  16\pi\, \tau^\m{}_I, \\\label{FET}
& P_{IJKL} \eps^{\m\n\r\s} e^K_\n T^L_{\r\s} = -16\pi\, \s^\m_{IJ}. 
\end{align}\end{subequations}
Here
\be\label{Gmi}
G^\m{}_I(e,\om)  := \f1{4}\eps_{IJKL}\eps^{\m\n\r\s} e_\n^J F_{\r\s}^{KL}(\om) = G^{\m\n}(e,\om)e_{\n I}
\ee
is the first-order Einstein tensor, the Riemann tensor and curvature are related by $R_{\m\n\r\s}(e,\om)=e_{\m I} e_{\n J}F^{IJ}_{\r\s}(\om)$, and $T^I:=d_\om e^I$ is the torsion.
The first set \Ref{FEE}  contains the ten Einstein equations, plus six redundant equations: although $G_{\m\n}(e,\om)$ is not symmetric a priori, 
it is easy to show that the Noether identity associated with invariance of the action under internal Lorentz transformations (see \Ref{N2} below)  implies that the equations for $G^\m{}_{[I}e_{J]\m}$ are automatically satisfied. The relevant content of \Ref{FEE} is therefore just its symmetric part, which in turn gives the Einstein's equations
\be
G_{\m\n}(e,\om) +\L g_{\m\n} +\f1{2\g } \eps_{(\m}{}^{\l\r\s} R_{\n)\l\r\s}(e,\om) =  8\pi\, T^{\G}_{\m\n}, \label{ECG}
\ee
or equivalently as functions of $(g,\G)$ via \Ref{gG}.

In the following, we will refer to \Ref{FEE} or \Ref{ECG} as Einstein's equations (in the presence of torsion), to be distinguished from the torsion Einstein-Cartan equations \Ref{FET}, or torsion equations for short.
It is often convenient to write the field equations using the language of differential forms, as we did in the action \Ref{SEC}. To that end, we use the
Hodge dual $\star$ mapping $p$-forms to $(4-p)$-forms (see Appendix~\ref{AppA} for conventions). 
This allows us to define the Einstein 3-from
\be
\star\!G_I(\om):= - \f12\eps_{IJKL}e^J\w F^{KL}(\om), 
\ee
where the opposite sign with respect to \Ref{Gmi} is a consequence of Lorentzian signature, and equivalently the dual source forms $\star\t_I$ and $\star\s_{IJ}$. The field equations \Ref{FE} then read
\begin{subequations}\label{FFE}\begin{align}
& \star \!G_I(\om) + \L \star \!e_I - \f1\g e^J\w F_{IJ}(\om) = 16\pi\,\star\! \tau_I, \\\label{FFET}
& P_{IJKL} \, e^K\w T^L = 8\pi\,\star\! \s_{IJ}.  
\end{align}\end{subequations}

\subsection{The contorsion tensor}
Although connections form an affine space with no preferred origin, the presence of an invertible tetrad suggests a natural origin: the Levi-Civita connection $\om_\m^{IJ}(e)$ associated with the tetrad. We can then always decompose an arbitrary connection into Levi-Civita plus a contorsion tensor $C_\m^{IJ}$ as 
\be\label{defC}
\om_\m^{IJ} = \om_\m^{IJ}(e) +C_\m^{IJ}. 
\ee
Torsion and curvature are related to the contorsion as follows:
\begin{align}
& T^I=C^{IJ}\w e_J, \\ \label{FdC}
& F^{JK}(\om) = F^{JK}(e) + d_{\om(e)}C^{JK} + C^{JM}\w C_M{}^K = F^{JK}(e) + d_{\om}C^{JK} - C^{JM}\w C_M{}^K,
\end{align}
where $d_{{\om(e)}} $ is the  exterior derivative with respect to the Levi-Civita connection. 
Plugging this decomposition into the field equations we find
\begin{align}\label{FEC1}
& \star \!G_I(e) + \L \star \!e_I = 16\pi\,\star\! \tau_I + P_{IJKL}( d_{\om(e)}C^{JK} + C^{JM}\w C_M{}^K), \\
& P_{IJKL} e^K\w C^{LM}\w e_M = 8\pi\, \star\! \s_{IJ}.\label{FEC2}
\end{align}

The fact that the field equations for the Einstein-Cartan theory can be recasted as in \Ref{FEC1} is the source of an old debate in the literature about the role of torsion \cite{hehl2007note}: if we forget about the notion of affine parallel transport defined by $\om^{IJ}$, and use simply the one defined by $\om^{IJ}(e)$ in the sector of invertible tetrads, then the theory is indistinguishable from ordinary metric theory with some non-minimal matter coupling. The non-minimality is captured by the effective energy-momentum tensor sourcing \Ref{FEC1}, i.e.
\be\label{teff}
\star\! \tau^{\scr eff}_I:=\star \tau_I + \f1{16\pi\,}P_{IJKL}( d_{\om(e)}C^{JK} + C^{JM}\w C_M{}^K).
\ee 
While we take no stand in the debate, we will heavily use this fact in the thermodynamic discussion below.
Before getting there, we need to review in the next Section the relation between the conservation of the energy-momentum tensor and the Bianchi identities.

For convenience of the reader, we report the relation between torsion and contorsion in tensor language, 
\begin{align}
& T^\r{}_{\m\n}:=e^\r_I \, T^I{}_{\m\n} = -2 C_{[\m,\n]}{}^\r=2\G^\r_{[\m\n]}, \\ & C_{\m,\n\r} = \f12 T_{\m,\n\r} - T_{[\n,\r]\m},\qquad C_{(\m,\n)\r} = T_{(\m,\n)\r}.
\end{align}
The Einstein equations \Ref{FEE} read
\begin{align}\label{FEEC}
& G_{\m\n}(e)+\L g_{\m\n} = 8\pi\, T^{\scr eff}_{\m\n}, \\
& T^{\rm eff}_{\m\n} = 2\t^I{}_{(\m} e_{\n)I} + \f1{16\pi\,}\Big(6g_{\a(\m}\d^{\a\r\s}_{\n)\g\d}-\f2\g g_{\g(\m}\eps_{\n)\d}{}^{\r\s}\Big) 
\Big(\og{\na}_\r C_{\s,}{}^{\g\d} + C_{\rho,}{}^{\gamma \lambda} C_{\s,\l}{}^\d\Big).\label{Teff}
\end{align}
We refrained from expanding the completely antisymmetric $\d^{\a\r\s}_{\n\g\d}$ since no useful simplification occurs.
Notice that for a given contorsion we have a 1-parameter family of conserved energy momentum tensors, labeled by the Immirzi parameter.
Finally, using the torsion equations, $T^{\rm eff}$ can be seen to be linear in the source tensor of the Einstein equations, and contain derivative and quadratic terms in the source tensor of the torsion equations.

\section{Noether identities and conservation laws}

The gravity action \Ref{SEC} is invariant under internal Lorentz transformations
\be\label{gauge}
\d_\l e^I =\l^I{}_J e^J, \qquad \d_\l\om^{IJ} = -d_\om \l^{IJ},
\ee
as well as diffeomorphisms,\footnote{Note that the Lie derivatives \Ref{diffeos} are not gauge-covariant objects. It is often convenient to consider the linear combination of transformations $L_\xi = \pounds_\xi + \d_{\om\lrcorner\xi}$ which is covariant.
}
\begin{subequations}\label{diffeos}\begin{align}
& \d_\xi e^I=\pounds_\xi e^I = de^I\lrcorner\xi+d(e^I\lrcorner\xi) = d_\om e^I\lrcorner\xi+d_\om(e^I\lrcorner\xi) - (\om^I{}_J\lrcorner\xi) e^J, \\
& \d_\xi \om^{IJ}= \pounds_\xi \om^{IJ} = d\om^{IJ}\lrcorner\xi+d(\om^{IJ}\lrcorner\xi)=F^{IJ}\lrcorner\xi+d_\om(\om^{IJ}\lrcorner\xi).
\end{align}\end{subequations}
Specializing the variation of the action \Ref{SEC} to \Ref{gauge} and \Ref{diffeos} respectively, and integrating by parts, one obtains the following Noether identities,\footnote{To obtain \Ref{NG1}, we used the identity \Ref{cyc1} below. For the reader's convenience, we report the identities also in the more common $\g$-less case,
\begin{align}
& \star\! G_{[I}\w e_{J]} = - \f12 \eps_{IJKL} e^K\w d_\om  T^L,\qquad
& d_\om \star\! G_I = -\f12\eps_{IJKL} T^J\w F^{KL}.
\end{align}
}
\begin{subequations}\label{NG}\begin{align}\label{NG1}
& P_{IJKL} e^K\w F^{LM} \w e_{M} = P_{IJKL} e^K\w d_\om T^L,\\
& d_\om (P_{IJKL} e^J\w F^{KL}) = P_{IJKL} T^J\w F^{KL}.
\end{align}\end{subequations}
These are nothing but contracted forms of the Bianchi identities $d_\om F^{IJ}=0$, $d_\om T^I=F^{IJ}\w e_J$.
Using the field equations \Ref{FFE} in \Ref{NG} one finds additional relations for the matter sources,
\begin{subequations}\label{NM}\begin{align}
\label{NM2} & d_\om\star\!\s_{IJ} = 2\star\!\t_{[I}\w e_{J]}, \\
\label{NM1} & d_\om\star\!\t_I = \f12 F^{JK}\lrcorner e_I\w \star\s_{JK} + T^{J}\lrcorner e_I \w\star\t_{J}.
\end{align}\end{subequations}
These matter Noether identities can also be derived without reference to the field equations  \Ref{FFE}: they follow from invariance of the matter action  \Ref{dSm} under \Ref{gauge} and \Ref{diffeos}, on-shell of the matter field equations.
See \cite{Hehl:1985vi,Hehl:1994ue,Barnich:2016rwk} for more details.

Recall now that, in the metric formalism, invariance of the matter Lagrangian under diffeomorphisms guarantees the conservation of the energy-momentum tensor,
\be
\d_\xi L_m = d(L_m\lrcorner \xi) \quad \Rightarrow \quad \na_\m T^{\m\n} = 0, 
\ee
on-shell of the matter field equations. 
In the first-order formalism with tetrads, the energy-momentum tensor does not appear immediately in the field equations: the closest object we have is the source $\t$ of the Einstein's equations \Ref{FEE}. This quantity is however not conserved, as we can see from \Ref{NM1}, whose right-hand side does not vanish on-shell.  
Nevertheless, although $\t$ is not conserved, it is easy to identity an effective energy-momentum tensor which is conserved, thanks to the contorsion decomposition \Ref{FEC1}. If we take the Levi-Civita  exterior derivative $d_{{\om(e)}} $  
on both sides of \Ref{FEC1}, the left-hand side vanishes identically. This in turns implies the vanishing of the right-hand side, which gives a local conservation law 
\be\label{dt0}
d_{{\om}(e)}\t^{\scr eff}_I = 0
\ee
valid \emph{also in the presence of torsion}. Equivalently in terms of tensors, the object with vanishing (Levi-Civita) divergence is $T^{\rm eff}_{\m\n}$ as defined in \Ref{Teff},
and it provides the conserved energy-momentum tensor of the theory.\footnote{An alternative `conservation law' using the full connection would be of little practical meaning, because it would not lead to hypersurface quantities independent of the choice of space-like slice.
Another way to identify this conserved object is to  
solve the torsion equation -- which in the case of Einstein-Cartan is simply algebraic since torsion does not propagate, and plug the solution back into the action. Varying the resulting matter action with respect to the tetrad then immediately gives the effective energy-momentum tensor \Ref{Teff}. }
This simple observation is well-known in the literature, see \cite{Hehl:1976kj,Hehl:1976vr,Boehmer:2017tvw} (where it is referred to as `combined energy-momentum tensor'), and can be taken to 
provide the basis of energy conservation in Einstein-Cartan theory.

For later purposes, we are interested in whether it is possible to derive the conservation law \Ref{dt0}
\emph{without} using the Einstein's equations. This is a bit of a strange question if one starts from an action principle, but it is crucial to Jacobson's thermodynamical argument, where this is not the case. We could not find the answer to this question in the literature, which turns out to be affirmative.
The result is the following:
\smallskip

{\bf Proposition 1:} \emph{
The matter Noether identities {\rm\Ref{NM}} on-shell of the matter and torsion field equations imply the conservation law for the effective energy-momentum tensor {\rm \Ref{dt0}.}
}
\smallskip

The proof is a somewhat lengthy exercise in algebraic identities, and we leave it to Appendix 2.
We also looked for a stronger result, namely whether \Ref{dt0} also holds without imposing the torsion equation, but we did not succeed. The proof in the Appendix \Ref{AppIndexjug} shows explicitly the step in which we use the torsion field equation. 
To give an idea of what happens, using the contorsion decomposition \Ref{NM} can be combined to give
\be\label{juantorena}
d_{\om(e)} \Big(\star\!\t_I +\f12 C^{JK}\lrcorner e_I  \star\!\s_{JK}\Big) =  \f12 \Big( F^{JK}(\om)\lrcorner e_I + \pounds_{e^I} C^{JK}\Big) \w \star \s_{JK}
\ee
(an expression for the Noether identities which appears for instance in \cite{Hehl:2013qga}), and using the torsion field equation the right-hand side reduces to $d_{\om(e)}$ of a 3-form.

\smallskip

In tensorial language, the Noether identities for a generic gauge and diff-invariant Lagrangian density $\cal L$ read (see e.g. \cite{Barnich:2016rwk})
\begin{subequations}\label{N}\begin{align}
\label{N2} & D_\m \f{\d \cL}{\d \om^{IJ}_\m} + \f{\d \cL}{\d e^{[I}_\m} e_{J]}^\m = 0,\\ 
\label{N1} & 
\f{\d \cL}{\d \om^{IJ}_\m} F^{IJ}_{\n\m}(\om)+ \f{\d \cL}{\d e^{I}_\m} T_{\n\m}^I - e_\n^I D_\m \f{\d \cL}{\d e^{I}_\m}  = 0, 
\end{align}\end{subequations}
on-shell of the matter field equations. For the Lagrangian density in \Ref{SEC}, these give respectively contractions of the algebraic and differential Bianchi identities,
\begin{align}
& \label{Bianchi2} 2R_{[\m\n]} = -\na_{\r} T^\r{}_{\m\n} -2\na_{[\m} T^\r{}_{\n]\r} + T^\r{}_{\r\s}T^\s{}_{\m\n}, 
\\ & \label{Bianchi1} 
\na_\n G^\n{}_\m = T^\r{}_{\m\s} R^\s{}_\r - \f12 T^\n{}_{\r\s} R^{\r\s}{}_{\m\n},
\end{align}
from the $\g$-less terms, and 
\begin{align}
& \eps^{\a\n\r\s}R_{\m\n\r\s} = \eps^{\a\n\r\s}(\na_\n T_{\m,\r\s} + T_{\m,\l\n} T^\l{}_{\r\s} ), \\
& \eps^{\a\b\r\s}\na_\b R_{\m\n\r\s} = \eps^{\a\b\r\s} T^\l{}_{\b\r} R_{\m\n\s\l}
\end{align}
for the part in $1/\g$. As for the matter action,
\begin{align}
& D_\m(e\s^\m_{IJ}) = -2e\t^\m{}_{[I}e_{J]\m},\\
& D_\m(e\t^\m{}_I) = ee^\m_I \left( \f12 F_{\m\n}^{JK}\s^\n_{JK} + T^J_{\m\n}\t^\n{}_J  \right).\label{Nm1}
\end{align}

\section{Einstein equations from thermodynamics}
We now come to the main motivation for our paper: show that Proposition 1 allows us to run Jacobson's argument with the usual equilibrium assumptions.
To better appreciate our point, let us briefly recall the key steps of the metric case, referring the reader to \cite{Jacobson:1995ab} for more details.

\subsection{The metric case}
Consider an arbitrary metric $g_{\m\n}$ on a manifold, a  point $P$ and a neighbourhood sufficiently small for spacetime to be approximately flat. 
Denote by $\xi^\m$ the future-pointing (approximate) Killing vector generating a Rindler horizon $\cal H$ within the approximately flat region, with bifurcating surface $\cal B$ through the point $P$. This is by construction hypersurface orthogonal, null at the horizon but not outside, and vanishing at $\cal B$:
\be
\xi^2\stackrel{\cal H}{=}0, \qquad \p_\m \xi^2 =:-2\, \k\, \xi_\m, \qquad \xi^\m\stackrel{\cal B}{=}0.
\ee
Since it is Killing, it is also geodesic, 
\be\label{geoLC}
\xi^\n\og{\na}_\n\xi^\m = - \f12 \p_\m\xi^2 = \k\, \xi^\m. 
\ee
The inaffinity $\k$ can be proven to be constant on the horizon, and it is usually referred to as the horizon surface gravity. For a Rindler horizon, constancy of $\k$ follows immediately from the vanishing of the Riemann tensor.\footnote{For a stationary black hole horizon, this is the content of the zeroth law of black hole mechanics. This was proved using the Einstein's equations and the dominant energy condition \cite{Bardeen:1973gs},
although in principle one could just require the analogue of the dominant energy condition directly on the Ricci tensor, as done in the generalization to isolated horizons \cite{Ashtekar:2000hw}.}
It is useful to introduce an affine parameter $\l$ along the null geodesics,  with origin at the point $P$. It can be easily shown that 
\be\label{xil}
\xi^\m=-\l \, \k \, l^\m, \qquad l^\m\p_\m = \p_\l.
\ee
\begin{figure}[h!]
\begin{center} \includegraphics[width=.25\textwidth]{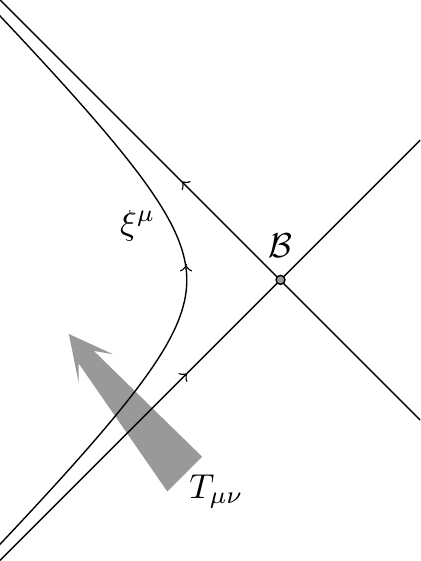} \end{center}
\caption{\small{\emph{The set-up thermodynamical derivation of Einstein's equation as proposed in \cite{Jacobson:1995ab}. Local flatness allows to consider approximate Rindler observers $\xi^\mu$ around any point $P$ of a given spacetime. The associate Rindler horizon has bifurcate surface $\mathcal{B}$ passing through $P$. The system is perturbed by a small flux of matter crossing the past horizon and entering the left wedge. For the derivation to be valid, an infinite family of $\xi^\mu$ is actually considered, one per each direction.}}}
\label{Fig} 
\end{figure}

Given this geometric set-up, the first step of Jacobson's argument is to associate to the Rindler horizon its Unruh temperature: 
\be\label{Tunruh}
(i) \qquad T=\f\k{2\pi}, \qquad \k={\rm constant}.
\ee
 Next, three assumptions are made: first, that there is an energy flux through the horizon in the near past of $P$, see Fig.~\ref{Fig}, given by  a \emph{conserved} energy-momentum tensor $T_{\m\n}$: 
 \be\label{defDU}
(ii) \qquad \D U := \int_{\cal H} T_{\m\n}\xi^\m l^\n d\l d^2S = - \k \int_{\cal H} T_{\m\n}l^\m l^\n \l d\l d^2S, \qquad \na_\m T^\m{}_\n = 0,
\ee
where  we used \Ref{xil} and the constancy of $\k$. 
This energy flux will be interpreted thermodynamically as a heat flux, $\D U=\D Q$. 
Second assumption, that there is a notion of entropy variation associated to the horizon, which is (universally, i.e. independently of the matter state) proportional to the area variation:
\be\label{iii}
(iii) \qquad \D S=\eta \,{\D A}  = \eta \int_{\cal H} \th d\l d^2S, 
\ee
where $\th$ is the expansion of horizon. This is controlled by the Raychadhuri equation for $l^\m$,
\be\label{Ray}
\f{d \theta}{d\l} = - \f{\theta^2}{2} - \sigma_{\mu \nu} \sigma^{\mu \nu} -R_{\mu \nu} l^\mu l^\nu.
\ee
The final, technical assumption made in \cite{Jacobson:1995ab} is that at $P$ one can take $\th=\s_{\m\n}=0$, and approximate the solution of the Raychadhuri equation simply by $\th=-\l R_{\mu \nu} l^\mu l^\nu+O(\l^2)$.\footnote{Vanishing of the initial expansion and shear are taken to be the equilibrium conditions necessary for the upcoming application of Clausius relation. We find on the other hand the last approximation quite strong in that it implies constant curvature, at least along the horizon's generators.  
See Appendix~\ref{AppTeodoro} for a discussion of this approximation, and an alternative derivation which uses perturbation theory in the metric fluctuations.} 
Using this approximation, 
\be\label{th95}
 \D S = -\eta \int_{\cal H}  \l\,R_{\m\n} l^\m l^\n d\l d^2S.
\ee

Finally, we observe that using $(i-iii)$ and the approximation \Ref{th95}, the Clausius first law of thermodynamics $\D Q=T\D S$ implies
\be\label{thermo}
\int_{\cal H} \left(\frac{2\pi}{\eta}\,T_{\m\n} -R_{\m\n}\right)l^\m l^\n \l d\l d^2S=0.
\ee
Since this is valid for an arbitrary direction of the Killing boost and at any point, we can remove the integral. The Einstein equations (with an undetermined cosmological constant) then follow by imposing the conservation law $\na^\m T_{\m\n}=0$. The Newton constant is identified determined by $G= 1/(4\eta)$.

\subsection{The torsional case}
In the Einstein-Cartan theory \Ref{SEC} the connection is a priori affine, and torsion can be present, affecting the geodesic and Raychaudhuri equations. One may then think that the argument above should be substantially revisited. As we now show, this is actually not the case.
The first observation we make is that the starting point of Jacobson's argument, a Killing horizon, is
a purely metric notion:
\begin{align}
0 = \pounds_\xi g_{\m\n} &=\xi^\a\p_\a g_{\m\n}+g_{\m\a}\p_{\n}\xi^\a+g_{\n\a}\p_\m\xi^\a & \\\nn
& = 2\og{\na}_{(\m}\xi_{\n)} = \na_{(\m} \xi_{\n)} + T_{(\m}{}^\r{}_{\n)} \xi_\r.
\end{align}
Hence by definition, it does not depend on torsion, in spite of the apparent presence of the latter in the last expression above. The constancy of $\k$ on the approximate Rindler horizon also follows like in the metric case from the vanishing of the metric Riemann tensor.
Being Killing and null, $\xi^\m$ is automatically geodetic with respect to the Levi-Civita connection (which we recall is always well-defined and at disposal since we are only interested in the sector of Einstein-Cartan theory with invertible tetrads), so \Ref{geoLC} still holds. Hence, we can run most of the argument as in the metric case.
Step $(i)$ is unchanged. 
For step $(ii)$, we follow \cite{Jacobson:1995ab} and define the energy flux as the integral of the \emph{conserved} energy-momentum tensor. Proposition 1 identifies this object uniquely as $T^{\rm eff}_{\m\n}$ defined in \Ref{Teff}, with its torsional dependence.
Step $(iii)$ is also unchanged: since 
the generators of the Killing horizon follow the Levi-Civita geodesics \Ref{geoLC},  
the change of the expansion of the generators is governed by the Raychaudhuri equation \Ref{Ray} with the metric Ricci tensor $R_{\m\n}(e)$ appearing on the right-hand side. Imposing again the equilibrium Clausius relation $\D Q=T \D S$ with these $(i-iii)$, and using the same approximation \Ref{th95}, we arrive exactly at
\be
\int_{\cal H} \left(\frac{2\pi}{\eta} \,T^{\rm eff}_{\m\n} -R_{\m\n}(e)\right)l^\m l^\n \l d\l d^2S.
\ee
We conclude that the torsion-full Einstein equations, in the form \Ref{FEEC}, can be derived \`a la Jacobson from the equilibrium Clausius relation. No need to consider a torsion-full Raychaudhuri equation, non-equilibrium terms and restrictions on torsion, as argued in \cite{Dey:2017fld} and reviewed in the next Section. It suffices to use the result of Proposition 1 to identify the correct energy-momentum tensor.

There is however an important caveat to our procedure: we are assuming the torsion equations to hold, since we used them to prove Proposition 1.
This may look unsatisfactory, since it is currently not known whether these equations can be derived from a thermodynamical description. 
Our logic is that if such a description of the torsion equations exists, then it is consistent to assume that they hold when deriving the Einstein equations. This said, it is also possible that Proposition 1 holds off-shell of the torsion equations, so that these are not needed to derive the Einstein equations.  Nonetheless, one would still need to be able to derive the torsion equations from thermodynamics for the whole framework to make sense. Assuming them to hold seems thus to us coherent if a complete thermodynamical framework exists.
In any case, the main problem if one does not want to use the conserved energy-momentum tensor is the ambiguity that one faces in defining it, see e.g. \cite{Hehl:1976vr}. 
The prescription used by the authors of \cite{Dey:2017fld} for instance, is to take what would be the source of the Einstein equations, namely the derivative of the matter Lagrangian with respect to the tetrad (or to the metric, equivalently up to a symmetrization). Notice that this can be tricky in the 
presence of torsion, because one can work with either the first-order action $S(g,\G)$ or the second-order action $S(g,C)$. The field equations are completely equivalent since the two actions are related by a (non-linear) field redefinition, however for the sources one has
\begin{align}\label{Tam}
& T^\G_{\m\n}:= \f 2{\sqrt{-g}}\f{\d L_m(g,\G)}{\d g^{\m\n}}, \\\label{Tam2}
& T^C_{\m\n} = \f 2{\sqrt{-g}}\f{\d L_m(g,C)}{\d g^{\m\n}}=T^\G_{\m\n} + \f 2{\sqrt{-g}}\f{\d L_m(g,\G)}{\d \G^\a_{\b\g}}\f{\d \G^\a_{\b\g}}{\d g^{\m\n}}.
\end{align}
Both coincide with the general relativity energy-momentum tensor when torsion vanishes, but differ in the presence of torsion. 
This type of ambiguity reminds us that using a conserved energy-momentum tensor, when available, is always the best choice.
We now show how this ambiguity in turn affects the non-equilibrium approach to the derivation of the Einstein equations.

\subsection{Non-equilibrium approach}

A more general setting including a non-vanishing shear has been considered in \cite{Eling:2006aw,Chirco:2009dc}. In this case the presence of additional terms on the right-most side of \Ref{iii} is incompatible with the equilibrium Clausius relation. Hence to run Jacobson's argument one must assume that there are non-equilibrium terms, 
\be\label{Cnonequi}
\D Q=T\D S+\D S_{\scr non-equi}.
\ee 
The interpretation of the shear-squared terms as non-equilibrium is justified a priori from the horizon tidal heating effect \cite{Chirco:2009dc}. 
It should be noticed however that the same shear-squared terms enter both the $T\D S$ and the $\Delta S_{\rm non-equi}$ contributions, since one is still assuming  \Ref{iii}, that the entropy variation is proportional to the area variation. This feature seems to us unusual from a thermodynamical perspective.

In any case, we now discuss the application of the non-equilibrium approach to deriving the Einstein equations, which is more problematic.
We start as before from the observation that a Killing horizon is metric-geodetic, and use the same approximations leading to the integrated metric Raychaudhuri equation \Ref{th95}, but this time allowing a non-zero shear in \Ref{Ray}. 
Then from \Ref{Cnonequi} we obtain
\be\label{thermo?}
 \f{2\pi}{\eta} \int_{\cal H}T^{\rm ??}_{\m\n} l^\m l^\n \l d\l d^2S= \int_{\cal H} \Big(R_{\m\n}(e) l^\m l^\n+\sigma_{\m\n}\s^{\m\n}\Big) \l d\l d^2S+ \D S_{\scr non-equi}.
\ee
The delicate point now is how to define the heat flux, namely what $T^{??}_{\m\n}$ needs to be used on the left-hand side of the above equation. Clearly, the identification of the non-equilibrium terms that will be needed to obtain the Einstein equations \Ref{FEEC} depends on how we define the energy-momentum tensor. If, as in the previous Section, the conserved one is used, the only non-equilibrium term comes from the shear, which can then be argued for as in the metric theory following \cite{Eling:2006aw,Chirco:2009dc}. 
This shows how the derivation of the Einstein equations from the conserved energy-momentum tensor and metric Raychaudhuri equation can be easily extended to the presence of shear.

If we chose instead to define the heat flux via a source tensor, we would need additional non-equilibrium terms in order to fully reproduce the Einstein equations \Ref{FEEC}. The crucial point is whether they can be justified a priori as in the example of the tidal heating, else the construction is artificial. 
The authors of \cite{Dey:2017fld} argue that this is possible, if $(a)$ we choose $T^{??}_{\m\n}=T^C_{\m\n}$ for the heat flux, and $(b)$ we define the non-equilibrium terms as those arising from the torsion-full Raychaudhuri equation that include torsion-full derivatives of $l^\m$. 
There are three problems that we can see with this construction. First, a Killing vector is metric-geodesic, but in general not geodesic with respect to the torsion-full connection, since from \Ref{defC} we see that
\be\label{geoaff}
\xi^\n{\na}_\n\xi_\m = \k\, \xi_\m - C_{\n,\m\r}\,\xi^\n\xi^\r = \k\, \xi_\m - T_{\n,\m\r}\,\xi^\n\xi^\r.
\ee
For this reason, the authors of \cite{Dey:2017fld} restrict torsion to satisfy
\be\label{Cll}
C_{\n,\m\r}\xi^\n\xi^\r = 0. 
\ee
This restriction implies that metric and torsion-full geodesics coincides, and one can use the geodesic torsion-full Raychaudhuri equation on the Killing horizon. But since the metric and the torsion-full geodesic expansions also coincide,\footnote{In the presence of torsion, the displacement of a vector $q^\m$  Lie dragged along $\xi^\m$ is given by $$
\xi^\n\na_\n q^\m = B_{\m\n}q^\n, \qquad B_{\m\n}:=\na_\n \xi_\m + T_{\m,\l\n}\xi^\l = \og\na_\n \xi_\m +C_{\r,\m\n}\xi^\r,
$$
hence introducing the usual projector $\perp^{\m\n}$ on a 2d space-like surface orthogonal to $\xi^\m$, we have \linebreak $\th:=\perp^{\m\n}B_{\m\n}=\og\th$. For the reader interested in more details on geodesics with torsion, see e.g. \cite{Luz:2017ldh,IoGT}.}
it follows that the torsion-full Raychaudhuri equation is \emph{identical} to the metric one. 
Therefore, it is unclear what one gains from this approach, except for a restriction on torsion that in the equilibrium approach presented in the previous Section is not necessary.\footnote{Since in order to recover the Einstein equations we will need to consider arbitrary boost Killing vectors, see discussion below \Ref{thermo}, the restriction on torsion \Ref{Cll} should hold for any $\xi^\m$. This implies a strong restriction on torsion, that can be satisfied for instance if it is completely antisymmetric. A priori it could be possible to consider a relaxation of \Ref{Cll}, allowing for a right-hand side proportional to $\xi_\m$ rather than vanishing, since this would only mismatch the inaffinity of metric and torsion-full geodesics. However we don't know whether the derivation of  \cite{Dey:2017fld} can be extended to this case.
}

Second, the identification of the non-equilibrium contributions as torsion-full covariant derivatives of $l^\m$ is questionable: 
we are not aware of any proof that in a spacetime with torsion it is the torsion-full shear that gives the tidal heating. Furthermore, the condition of vanishing initial expansion implies that at the point $P$ we have $\na_\m l^\m= \perp^{\m\n}\!T_{\m,\n\r}l^\r$, making some `non-equilibrium terms' indistinguishable from terms without derivatives, as the authors of  \cite{Dey:2017fld} acknowledge in a footnote. 

Third, there is the ambiguity associated with picking a non-conserved $T^{??}_{\m\n}$, as discussed before. Had we chosen the alternative source $T^\G$, which is also more natural from the perspective of a metric-connection action, the same identification of non-equilibrium contributions would not work, as it would miss the terms with covariant derivatives of the contorsion in \Ref{FEEC}.

Summarizing, although the non-equilibrium approach has the advantage of allowing to relax the assumption of an initial non-vanishing shear \cite{Eling:2006aw,Chirco:2009dc}, 
it is in our opinion ambiguous when applied to gravity with torsion.

\section{On the laws of black hole mechanics with torsion}\label{SectionLaws}

As mentioned in the introduction, Jacobson's derivation is inspired by the laws of black hole thermodynamics. Having shown that the derivation works also in the presence of torsion, at least as far as recovering the Einstein equations, the next question we considered is what happens to the these laws. 

We have recalled earlier that the surface gravity of the Rindler horizon is constant simply because the Riemann tensor vanishes. For a general horizon, constancy of the surface gravity is the zeroth law, and its proof uses the Einstein equations and the dominant energy conditions.
In the presence of torsion, we can follow the proof with the equations \Ref{FEEC}, and the only modification is that the dominant energy condition will be a restriction on the effective energy-momentum tensor.

More interesting is the modification that occurs to the first law. 
To see this, let us consider the `physical process' version of the proof \cite{Wald:1995yp}, in which an initially stationary black hole is perturbed by some matter falling inside the horizon. For our generalization, we suppose that the in-falling matter has spin and sources torsion, and that the metric and connection satisfy the Einstein-Cartan field equations.

As in the metric case, we assume  that all matter falls into the black hole, and that the latter is not destroyed by the process, but settles down to a new stationary configuration 
\cite{Wald:1995yp,Gao:2001ut}. These assumptions are motivated by the no-hair theorem and the cosmic censorship conjecture, which keep their value also in a theory with non-propagating torsion.
For example, it is known that a compact ball of static or slowly spinning torsion-full Weyssenhoff fluid\footnote{This is a single component of torsion (the trace part) generated by the gradient of a scalar \cite{griffiths1982spin}.} admits a solution which satisfies the junction conditions with an external Schwarzschild or slowly rotating Kerr \cite{Prasanna:1975wx,arkuszewski1974linearized}.

Following \cite{Wald:1995yp}, we use the linearized Einstein equation to study the effect on the horizon geometry caused by the in-falling matter at first order in perturbation theory,
\be
g_{\m\n}=g_{\m\n}^0+h_{\m\n}, \qquad C_{\r,\m\n}=c_{\r,\m\n}.
\ee 
Being null and hypersurface orthogonal, the affine horizon generators are metric geodetic, and their expansion is governed by the Raychaudhuri equation \Ref{Ray}. The background generators $l^\m$ are proportional to the Killing generators $\xi^\m$ satisfying $l^\m=-(\l\k)^{-1}\xi^\m$, with constant $\k$ by the zeroth law. They have vanishing shear and expansion, giving therefore at first order  
\be\label{eq:rayFL}
\f{d}{d\l} \delta\theta = -\delta R_{\mu \nu}(h) l^\mu l^\nu. 
\ee
Integrating along the horizon $\cal H$  from the bifurcation surface $\cal B$ to a cut $S_\infty$ at future null infinity, we have for the total area variation
\be\label{DA}
\D A = \int_{\cal H} \d\th \, d\l d^2S 
= \int_{\cal H} \delta R_{\mu \nu}(h) l^\mu l^\nu \, \l d\l d^2S,
\ee
where we integrated by parts and used that $\l|_{\cal B}=0$ since $\xi^\mu|_{\cal B}=0$, and that $\th|_{S_\infty}=0$ by the late time settling down assumption. 

In the standard particular case of torsion-less matter with conserved energy-momentum tensor $T_{\m\n}$, we have from the linearized Einstein equations 
\be
\int_{\cal H} \delta R_{\mu \nu}(h) l^\mu l^\nu \, \l d\l d^2S = 8\pi \int_{\cal H} \delta T_{\mu \nu}(h) l^\mu l^\nu \, \l d\l d^2S. 
\ee
At this order, we can substitute $l^\m=-(\l\k)^{-1}\xi^\m$ in the right-hand side integrand 
\be
- \f{8\pi}\k \int_{\cal H} \delta T_{\mu \nu}(h) \xi^\mu l^\nu \, \l d\l d^2S = \f{8\pi}\k \int_{\cal H} \delta T_{\mu\n}(h) \xi^\mu \, d H^\n =
\f{8\pi}\k ( \D M -\Omega_H\D J),
\ee
where in the first equality we used that fact the future-pointing volume form on $\cal H$ is \linebreak $d H_\m=-l_\m d\l d^2S$, and in the second the explicit expression $\xi^\m=\p_t^\m+\Omega_H\p_\phi^\m$ as well as the definitions of $\D M$ and $\D J$ used in \cite{Wald:1995yp}.
 We conclude that the linearized Einstein equations imply the first law of perturbations around a stationary black hole,\footnote{To make contact between this `physical process' version of the first law, and the one in terms of ADM (Arnowitt-Deser-Misner) charges, recall that since we are assuming all matter to be falling in the black hole, the integral along the horizon equals the integral on a space-like hypersurface $\Sigma$ extending from $\cal B$ to a 2-sphere $S_\infty$ at spatial infinity $i^0$. Using again the Einstein equations and the explicit form of the conserved Noether current (see \cite{Iyer:1994ys}, here $\k$ is the Komar charge and $\Theta$ the Einstein-Hilbert symplectic potential) we find
\be\nn
\int_{\cal H} \delta T_{\mu \nu}(h) \xi^\mu d H^\n = \int_{\Sigma} \delta T_{\mu \nu}(h) \xi^\mu d\Sigma^\n =
\int_{\cal S_\infty} (k_\xi-\Theta\lrcorner\xi) - \int_{\cal B} k_\xi= \D M_{ADM} - \Omega_H\D J_{ADM},
\ee
 where 
 the final result follows from a standard calculation with $\xi^\m=\p_t^\m+\Omega_H\p_\phi^\m$. See \cite{DePaoli:2018erh} for a derivation of the first law with covariant Hamiltonian methods for Einstein-Cartan theory.
 We remark that while completing our paper, similar considerations on the role of torsion in the first law appeared in \cite{Chakraborty:2018qew}, albeit with what seems to us a particular matter Lagrangian.}
 \be
 \D M = \f\k{8\pi}\D A + \Omega_H\D J.
 \ee
 
For torsion-generating matter, we can follow exactly the same procedure, the only difference being that we use the Einstein-Cartan equations \Ref{FEEC} with the conserved effective energy-momentum tensor on the right-hand side. The first law follows as before but with new mass and angular momentum variations
\be
 \D M -\Omega_H\D J = \int_{\cal H} \delta T^{\rm eff}_{\mu\n}(h) \xi^\mu \, d H^\n 
\ee
determined by the torsion-dependent $T^{\rm eff}_{\m\n}$.
This is consistent with the results of \cite{arkuszewski1974linearized} mentioned above, where the mass of the external Schwarzschild has a torsion contribution from an effective energy density profile of the static Weyssenhoff fluid compatible with the formula above.

Following the same approach of treating the effect of torsion as an effective energy-momentum tensor, we can conclude that also the second law of black hole mechanics is still valid, provided the required restrictions on the energy-momentum tensor of matter \cite{Bardeen:1973gs} are applied to the effective tensor \Ref{Teff}.

As for the more elusive third law, a discussion would require a prior understanding of extremal black holes in the presence of torsion, which lies beyond the scope of this paper.

\section{Conclusions}
Prompted by the analysis of \cite{Dey:2017fld}, we looked at one aspect of conservation laws in Einstein-Cartan theory. In the sector of invertible tetrads, where one can choose to split the connection into the Levi-Civita one plus a contorsion tensor, it is immediate to identify a conserved energy-momentum tensor $T^{\rm eff}$ for matter from the Einstein equations.  We showed in our paper that $T^{\rm eff}$ can be derived \emph{without} using the Einstein equations, starting instead from the Noether identities associated with the gauge and diffeomorphism invariance of the matter Lagrangian, and relating them through the torsion equations.

Thanks to this result, we were able to reproduce Jacobson's thermodynamical argument \cite{Jacobson:1995ab}, and derive the Einstein equations from the equilibrium Clausius relation. 
Our derivation is much simpler than the one proposed in \cite{Dey:2017fld}, and does not require non-equilibrium terms nor any restriction on torsion. On the other hand, like in \cite{Dey:2017fld}, we are only able to derive the tetrad Einstein equations from a thermodynamical argument, and not the torsion equations as well. This remains an open question in order to truly extend Jacobson's argument to theories with independent metric and connection.

For our construction, we used first the equilibrium set-up of \cite{Jacobson:1995ab}, in particular the initial shear vanishes.
Non-equilibrium terms have been advocated in order to relax this assumption  \cite{Eling:2006aw,Chirco:2009dc,Guedens:2011dy}, and the same can be done in the presence of torsion: we showed that one can treat the shear alone as non-equilibrium, and still derive the torsion-full Einstein equations with all the torsional dependence coming from the equilibrium part.

On the other hand, non-equilibrium terms could become crucial if one were able to go beyond Einstein-Cartan theory, and apply a thermodynamical reasoning to derive the field equations of modified theories of tetrad and connection with higher order terms, which typically include propagating torsion (and associated ghosts, see e.g. \cite{Tseytlin:1981nu}). 
It could be interesting if the dissipation present in this case would be associated with dissipation of energy to the torsional degrees of freedom. From this perspective, as well as the perspective of possibly recovering the torsion field equations from a thermodynamical argument, it could be intriguing to consider existing condensed matter models in which dissipating  lattice defects introduce torsion \cite{Dislocation}.

\subsection*{Acknowledgments}
We would like to thank the authors of \cite{Dey:2017fld} for discussions and comments on a draft of this paper.
TDL thanks Alejandro Perez for discussions.

\appendix
\setcounter{equation}{0}
\renewcommand{\theequation}{\Alph{section}.\arabic{equation}}

\section{Conventions}\label{AppA}

We take $\ut{\eps}_{\m\n\r\s}$ as the completely antisymmetric spacetime density with $\ut{\eps}_{0123}=1$, and 
$\tl\eps^{\m\n\r\s}\ut{\eps}_{\m\n\r\s}=-4!$. It is related to the volume 4-form by
\be
\eps:=\f1{4!}\eps_{\m\n\r\s}dx^\m\w dx^\n\w dx^\r\w dx^\s, \qquad \eps_{\m\n\r\s}:=\sqrt{-g} \, \ut{\eps}_{\m\n\r\s}.
\ee
We define the Hodge dual in components as
\be
(\star \om^{(p)})_{\m_1..\m_{4-p}} := \f{1}{p!} \om^{(p)}{}^{\a_1..\a_p} \eps_{\a_1..\a_p\m_{1}..\m_{4-p}}.
\ee

For the internal Levi-Civita density $\eps_{IJKL}$ we refrain from adding the tilde. We use the same convention, ${\eps}_{0123}=1$, so the tetrad determinant is
\be\label{tetId2}
e = -\f1{4!}\eps_{IJKL}\tl\eps^{\m\n\r\s} e_\m^I e_\n^J e_\r^K e_\s^L, 
\ee
and we take $e>0$ for a right-handed tetrad. 

Curvature and torsion are defined by
\be
F^{IJ}(\om) = d \om^{IJ} + \om^{IK}\w \om_K{}^J, \qquad \label{defT}
T^I(e,\om) = d_\om e^I,
\ee
where $d_\om$ is the covariant exterior derivative, whose components we denote by $D_\m$, to distinguish them from the spacetime covariant derivative $\na_\m$ with affine connection $\G^\r_{\m\n}$. The relation between the connections on the fiber and on the tangent space is given by
\be
D_\m e_\n^I = \G^\r_{\m\n} e_\r^I, \qquad \om^{IJ}_\m=e_\n^I{\na}_\m e^{\n J}
\ee
for $\om$ and $\G$ general affine connections, plus
the metricity condition $D_\m \eta^{IJ}=0$. The compatibility of the internal covariant derivative and the tetrad means that
$D_\m f^I=e^I_\n\na_\m f^\n$ and so on.

The commutators of the covariant derivatives satisfy: 
\begin{align}
& [D_\m,D_\n]f^I = F^I{}_{J\m\n}(\om)f^J, \\
& [D_\m,D_\n]f = - T^\r{}_{\m\n}(e,\om) \p_\r f, \\
& [\na_\m,\na_\n]f^\r = R^\r{}_{\s\m\n}(\G)f^\s - T^\s{}_{\m\n}\na_\s f^\r,
\end{align}
where
\begin{align}
R_{\r\s\m\n}(\G) = e_{I\r}e_{J\s} F^{IJ}_{\m\n}(\om) \qquad
T^\r{}_{\m\n}(\G) =e^\r_I \, T^I{}_{\m\n}(\om).
\end{align}
Finally, torsion and contorsion are related by 
\begin{align}
& T^\r{}_{\m\n}:=e^\r_I \, T^I{}_{\m\n}(e,C) = -2 C_{[\m,\n]}{}^\r=2\G^\r_{[\m\n]} 
\quad \Leftrightarrow \quad C_{\m,\n\r} = \f12 T_{\m,\n\r} - T_{[\n,\r]\m}. \label{CofT}
\end{align}
Both torsion and contorsion have spinorial decomposition
$
{\bf (\tfrac32,\tfrac12)\oplus(\tfrac12,\tfrac32)\oplus(\tfrac12,\tfrac12)\oplus(\tfrac12,\tfrac12)},
$
which corresponds to three irreducible components under Lorentz transformations (since the latter include parity). They can be defined as follows \cite{Hehl:1976kj},
\begin{align}
& C^{\m,\n\r} = \bar{C}^{\m,\n\r} + \f23 g^{\m[\rho} \check{C}^{\n]} + \eps^{\m\n\r\s} \hat C_{\s}, \\
& g_{\m\n} \bar{C}^{\m,\n\r}=0=\eps_{\m\n\r\s} \bar{C}^{\m,\n\r}, \qquad \check C^\m:=C_{\n,}{}^{\m\n}, \qquad \hat C_\s:=\f16\eps_{\s\m\n\r}C^{\m,\n\r}.
\end{align}

\section{Index jugglers}\label{AppIndexjug}
In this Appendix we prove Proposition 1, namely that the matter Noether identities \Ref{NM} on-shell of the matter field equations, plus the torsion field equation \Ref{FFET}, imply the conservation law for the effective energy-momentum tensor \Ref{dt0}, reported here for convenience
\be
\label{BI1}
d_{\om(e)}\left[\star\tau_I+\f1{16\pi\,}P_{IJKL} (e^J\w d_{\om(e)} C^{KL}+ e^J\w C^{KM} \w C_M{}^L)\right] = 0,
\ee
namely,
\be\label{BI2}
d_{\om(e)} \star\!\tau_I = \f1{8\pi\,}P_{IJKL} \Big(e^J\w C^K{}_M\w F^{ML}(e)+ e^J\w d_{\om(e)} C^{KM} \w C_M{}^L\Big).
\ee

To prove this identity, we start from \Ref{NM1}. On the left-hand side, we split the connection into Levi-Civita plus contorsion, see \Ref{defC}, obtaining
\be\label{LHS1}
d_\om\star\!\t_I = d_{\om(e)}\star\!\t_I-(C^{JK}\lrcorner e_I)e_K \w \star\t_J + T^J\lrcorner e_I \w \star\t_J
\ee
where we used
\be \label{tc}
T^I=C^{IJ}\w e_J\quad \rightarrow \quad C_I{}^{J}= -(C^{JK}\lrcorner e_I)e_K + T^J\lrcorner e_I.
\ee
In the second term of the right-hand side of \Ref{LHS1} we use the second Noether identity \Ref{NM2}, whereas  
the last term cancels the corresponding one on the right-hand side of \Ref{NM1}, which then reads
\begin{align}
d_{\om(e)} \star\!\t_I &=  \f12 F^{JK}(\om)\lrcorner e_I\w \star \s_{JK}-\f12 (C^{JK}\lrcorner e_I)d_\om \star\!\s_{JK} \nn\\
&=  \f1{16\pi\,} \left[F^{JK}(\om)\lrcorner e_I\w - (C^{JK}\lrcorner e_I)d_\om \right]P_{JKLM} e^L\w C^{M}{}_{N}\w e^N \nn \\
& = \f1{16\pi\,} \left[\big(F^{JK}(e)+ d_{\om(e)}C^{JK}\big)\lrcorner e_I\w - (C^{JK}\lrcorner e_I) d_{{\om}(e)} \right] P_{JKLM} e^L\w C^{M}{}_{N}\w e^N.
\label{BI3} 
\end{align}
In the second equality above we eliminated the torsion source using the corresponding field equation \Ref{FEC2}.
In the third equality we expanded the curvature using the contorsion, see \Ref{FdC}, and observed that the piece quadratic in $C$ cancels the contorsion part of the exterior derivative in the last term.\footnote{Following the same steps but without eliminating the torsion source in favour of the contorsion one gets \Ref{juantorena} in the main text, which does not use the torsion field equations but only the Noether identities.}
 
Having performed these simplifications, our goal is to show the equivalence of the right-hand sides of \Ref{BI2} and \Ref{BI3}. This will  follow from the equivalence of the terms with the Riemann tensor $F^{IJ}(e)$, and the equivalence of the terms involving the Levi-Civita exterior derivatives. Both are consequences of trivial algebraic symmetries. Let us show them one by one. We notice in advance the following useful cycling identities:
\begin{align}
& P_{IJKL} e^K\w F^{LM}\w e_M = -P_{ABC[I}e^A\w F^{BC}\w e_{J]}, \label{cyc1} \\
& P_{IJKL}  F^{KM}\w C_{M}{}^L = -P_{ABC[I}F^{AB}\w C^C{}_{J]}, \label{cyc2}
\end{align}
which are easy to check.

To show the equivalence of the terms with the curvature, we start hooking a cotetrad vector field on a trivially vanishing 5-form,
\begin{align}\label{5form}
0 &= \Big(P_{JKLM }F^{JK}(e)\w e^L \w C^{M}{}_{N} \w e^N\Big)\lrcorner e_I \nn\\
& = P_{JKLM}F^{JK}(e)\lrcorner e_I\w e^L \w C^{M}{}_N \w e^N + P_{JKIM }F^{JK}(e) \w C^M{}_N \w e^N 
\nn\\&\quad - P_{JKLM } (C^{M}{}_N \lrcorner e_I) F^{JK}(e)\w e^L \w e^N + P_{JKLM }F^{JK}(e) \w e^L \w C^{M}{}_I.
\end{align}
Of these four terms, the third vanishes identically: its $1/\g$ part directly through the algebraic Bianchi identities for the Riemann tensor, the other part because of the antisymmetry in the $LP$ indices. The second and fourth terms recombine giving the left-hand side of \Ref{cyc2}, hence \Ref{5form} gives
 \be
2 P_{IJKL}\, F^{KM}(e)\w C_M{}^L\w e^J = P_{JKML} \,(F^{JK}(e)\lrcorner e_I)\w e^L \w C^M{}_N \w e^N,
\ee
which proves the equality of the curvature terms of \Ref{BI2} and \Ref{BI3}.

The equivalence of the $d_{\om(e)}C$ terms follows analogously. We hook the following 5-form,
\begin{align}\label{5form2}
0 &= \Big(P_{JKLM }C^{JK}\w e^L \w d_{\om(e)} C^{M}{}_{N} \w e^N\Big)\lrcorner e_I \nn\\
& = P_{JKLM} ( C^{JK}\lrcorner e_I) e^L \w d_{\om(e)} C^{M}{}_N \w e^N - P_{JKIM }C^{JK} \w d_{\om(e)} C^M{}_N \w e^N 
\nn\\&\quad + P_{JKLM } C^{JK}\w e^L\w d_{\om(e)} C^{M}{}_N \lrcorner e_I \w e^N + P_{JKLM }C^{JK} \w e^L \w d_{\om(e)}C^{M}{}_I.
\end{align}
Using an identity like \Ref{cyc2}, the second and fourth term give
\be
P_{JKM[I }C^{JK} \w d_{\om(e)}C^{M}{}_{N]} \w e^N = -2 P_{IJKL }\w e^J \w d_{\om(e)}C^{K}{}_M \w C^{ML}. 
\ee
For the third term we have 
\begin{align}
& P_{JKLM } C^{JK}\w e^L\w d_{\om(e)} C^{M}{}_N \lrcorner e_I \w e^N = -P_{JKLM} d_{\om(e)} C^{MN}\lrcorner e_I \w e^L \w C^{JK} \w e_N \\\nn
& \qquad = P_{JKLM} d_{\om(e)} C^{JK}\lrcorner e_I \w e^L \w C^{M}{}_N \w e^N,
\end{align}
which follows from a similar cycling identity as before.
Hence, \Ref{5form2} gives
\begin{align}
2 P_{IJKL }\w e^J \w d_{\om(e)}C^{K}{}_M \w C^{ML} &=  P_{JKLM} d_{\om(e)} C^{JK}\lrcorner e_I \w e^L \w C^{M}{}_N \w e^N \\\nn
&\quad + P_{JKLM} ( C^{JK}\lrcorner e_I) e^L \w d_{\om(e)} C^{M}{}_N \w e^N,
\end{align}
which proves precisely the equivalence between the $d_{\om(e)}C$ terms in \Ref{BI2} and \Ref{BI3}.

\section{Horizon heat flux}\label{AppTeodoro}
\begin{figure}[h!]
\begin{center} \includegraphics[width=0.25\textwidth]{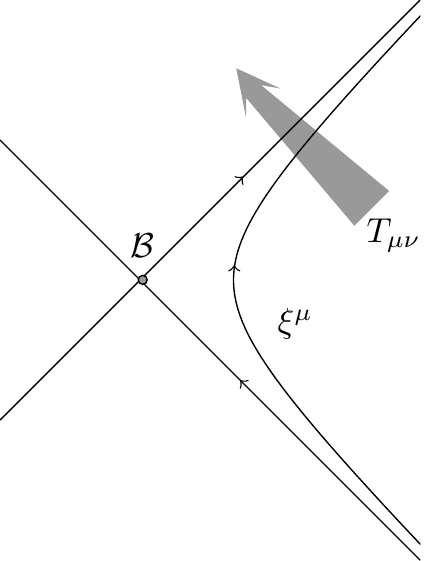} \end{center}
\caption{\label{Fig2} \small{\emph{The set-up thermodynamical derivation of Einstein's equation as proposed in \cite{Guedens:2011dy}. Local flatness allows to consider approximate Rindler observers $\xi^\mu$ around any point $P$ of a given spacetime. The associate Rindler horizon has bifurcate surface $\mathcal{B}$ passing through $P$. The system is perturbed by a small flux of matter crossing the future horizon and leaving the right wedge. An infinite family of $\xi^\mu$ is actually considered, one per each direction.}} }
\label{fig:11}
\end{figure}

In this Appendix we discuss some details of Jacobson's thermodynamic argument, and consider a different derivation motivated by the (backwards) similitude with the physical process proof of the first law of black hole thermodynamics.
Let us first review the physical set-up and its thermodynamical interpretation. With reference to Fig.~\ref{Fig}, we see that from the perspective of the boosted observer the energy flux is coming out of its `white hole horizon', or as the authors of \cite{Guedens:2011dy} put it, `one has to think of the heat as going into a reservoir which is behind the horizon'. We suppose this must be the reason why \Ref{defDU} is defined with a minus signs with respect to the outgoing energy flux (the future-pointing integration is $d{H}_\m = -l_\m d\l d^2S$, as we used in Section~\ref{SectionLaws}). 
An alternative set-up was presented in \cite{Guedens:2011dy}, see Fig.~\ref{fig:11}, placing the energy flux in the future of the bifurcation surface, so to have the boosted observer seeing it falling into its Rindler horizon.
Spacetime is initially flat, in particular $\th=\s_{\m\n}=0$ at the bifurcation surface. With the same approximations used in \Ref{th95} (i.e. constant curvature at first order in the affine parameter from $\cal B$), one can again derive the Einstein equation from the Clausius relation. 
 The physical interpretation of the Clausius law is the same: there is a negative energy flux which corresponds to a reduction in entropy, and assuming an entropy universally proportional to the area this translates into the focusing of geodesics. But now the initial heat reservoir is within the domain of causality of the boosted observer, which is an appreciable feature to have.
 
 Within this '11 set-up, the analogy between the argument and the first law is manifest, and it suggests an alternative procedure, with the advantage of relaxing the constant curvature approximation, at the price of an additional assumption. 
Consider the same set-up of Fig.~\ref{fig:11}, but let us assume this time that spacetime is initially arbitrary, and that long enough after the flux has crossed and perturbed the horizon, the latter `settles down' to Rindler again. This is an assumption, which in the case of the first law is motivated by the no-hair theorem; it has no corresponding backing-up in the case of a Rindler horizon that we know of, but we observe that the same assumption is used to derive the results of \cite{Bianchi:2013rya,DeLorenzo:2017tgx}. We can then treat the Raychaudhuri equation not at first order in $l$, which implies a constant curvature, but at first order in the metric perturbations, with small but otherwise arbitrary curvature along the horizon. Thanks to the assumption of Rindler behavior at later times we can obtain the area variation integrating by parts as in \Ref{DA} in the main text, without needing to know the explicit solution to the Raychaudhuri equation.

Then, using the same steps $(i-iii)$ (with a small energy-momentum tensor $\d T_{\m\n}$), but replacing \Ref{th95} with \Ref{DA}, we can again derive the Einstein equations.
This alternative derivation has the nice feature, to our taste, of not requiring constant curvature and energy-momentum tensors, with the consequence of making all $d\l$ integrations really not significative. However it can hardly be considered a more solid derivation, as we initially hoped, because of the ad hoc `Rindler stationarity' assumption at late times.
This could be removed if we reverse the boundary conditions, and required that spacetime is initially Rindler, namely at $\cal B$, and can be arbitrary at later times. However the derivation does not work unfortunately, unless curvature is constant again, which is what allows the authors of \cite{Guedens:2011dy} to reverse boundary conditions with respect to \cite{Jacobson:1995ab}.

\providecommand{\href}[2]{#2}\begingroup\raggedright\endgroup


\begin{thebibliography}{10}

\bibitem{Jacobson:1995ab}
T.~Jacobson, {\it {Thermodynamics of space-time: The Einstein equation of
  state}},  Phys. Rev. Lett. {\bf 75} (1995) 1260--1263
  [\href{http://arXiv.org/abs/gr-qc/9504004}{{\tt gr-qc/9504004}}].

\bibitem{Bardeen:1973gs}
J.~M. Bardeen, B.~Carter and S.~W. Hawking, {\it {The Four laws of black hole
  mechanics}},  Commun. Math. Phys. {\bf 31} (1973) 161--170.

\bibitem{Eling:2006aw}
C.~Eling, R.~Guedens and T.~Jacobson, {\it {Non-equilibrium thermodynamics of
  spacetime}},  Phys. Rev. Lett. {\bf 96} (2006) 121301
  [\href{http://arXiv.org/abs/gr-qc/0602001}{{\tt gr-qc/0602001}}].

\bibitem{Chirco:2009dc}
G.~Chirco and S.~Liberati, {\it {Non-equilibrium Thermodynamics of Spacetime:
  The Role of Gravitational Dissipation}},  Phys. Rev. {\bf D81} (2010) 024016
  [\href{http://arXiv.org/abs/0909.4194}{{\tt 0909.4194}}].

\bibitem{Guedens:2011dy}
R.~Guedens, T.~Jacobson and S.~Sarkar, {\it {Horizon entropy and higher
  curvature equations of state}},  Phys. Rev. {\bf D85} (2012) 064017
  [\href{http://arXiv.org/abs/1112.6215}{{\tt 1112.6215}}].

\bibitem{Dey:2017fld}
R.~Dey, S.~Liberati and D.~Pranzetti, {\it {Spacetime thermodynamics in the
  presence of torsion}},  Phys. Rev. {\bf D96} (2017), no.~12 124032
  [\href{http://arXiv.org/abs/1709.04031}{{\tt 1709.04031}}].

\bibitem{Wald:1995yp}
R.~M. Wald, {\em {Quantum Field Theory in Curved Space-Time and Black Hole
  Thermodynamics}}.
\newblock Chicago Lectures in Physics. University of Chicago Press, Chicago,
  IL, 1995.

\bibitem{Hehl:1994ue}
F.~W. Hehl, J.~D. McCrea, E.~W. Mielke and Y.~Ne'eman, {\it {Metric affine
  gauge theory of gravity: Field equations, Noether identities, world spinors,
  and breaking of dilation invariance}},  Phys. Rept. {\bf 258} (1995) 1--171
  [\href{http://arXiv.org/abs/gr-qc/9402012}{{\tt gr-qc/9402012}}].

\bibitem{hehl2007note}
F.~W. Hehl and S.~Weinberg, {\it Note on the torsion tensor},  Physics Today
  {\bf 60} (2007), no.~3 16.

\bibitem{Hehl:1985vi}
F.~W. Hehl and J.~D. McCrea, {\it {Bianchi Identities and the Automatic
  Conservation of Energy Momentum and Angular Momentum in General Relativistic
  Field Theories}},  Found. Phys. {\bf 16} (1986) 267--293.

\bibitem{Barnich:2016rwk}
G.~Barnich, P.~Mao and R.~Ruzziconi, {\it {Conserved currents in the Cartan
  formulation of general relativity}},
  [\href{http://arXiv.org/abs/1611.01777}{{\tt 1611.01777}}].

\bibitem{Hehl:1976vr}
F.~W. Hehl, {\it {On the Energy Tensor of Spinning Massive Matter in Classical
  Field Theory and General Relativity}},  Rept. Math. Phys. {\bf 9} (1976)
  55--82.

\bibitem{Boehmer:2017tvw}
C.~G. B{\"o}hmer and F.~W. Hehl, {\it {Freud's superpotential in general
  relativity and in Einstein-Cartan theory}},  Phys. Rev. {\bf D97} (2018),
  no.~4 044028 [\href{http://arXiv.org/abs/1712.03268}{{\tt 1712.03268}}].

\bibitem{Ashtekar:2000hw}
A.~Ashtekar, S.~Fairhurst and B.~Krishnan, {\it {Isolated horizons: Hamiltonian
  evolution and the first law}},  Phys. Rev. {\bf D62} (2000) 104025
  [\href{http://arXiv.org/abs/gr-qc/0005083}{{\tt gr-qc/0005083}}].

\bibitem{Luz:2017ldh}
P.~Luz and V.~Vitagliano, {\it {Raychaudhuri equation in spacetimes with
  torsion}},  Phys. Rev. {\bf D96} (2017), no.~2 024021
  [\href{http://arXiv.org/abs/1709.07261}{{\tt 1709.07261}}].

\bibitem{IoGT}
S.~Speziale, {\it {Raychaudhuri and optical equations for NGCs with torsion}},
  [\href{http://arXiv.org/abs/1808.00952}{{\tt
  1808.00952}}].

\bibitem{Gao:2001ut}
S.~Gao and R.~M. Wald, {\it {The 'Physical process' version of the first law
  and the generalized second law for charged and rotating black holes}},  Phys.
  Rev. {\bf D64} (2001) 084020 [\href{http://arXiv.org/abs/gr-qc/0106071}{{\tt
  gr-qc/0106071}}].

\bibitem{griffiths1982spin}
J.~Griffiths and S.~Jogia, {\it A spin-coefficient approach to weyssenhoff
  fluids in einstein-cartan theory},  General Relativity and Gravitation {\bf
  14} (1982), no.~2 137--149.

\bibitem{Prasanna:1975wx}
A.~R. Prasanna, {\it {Static Fluid Spheres in Einstein-Cartan Theory}},  Phys.
  Rev. {\bf D11} (1975) 2076--2082.

\bibitem{arkuszewski1974linearized}
W.~Arkuszewski, W.~Kopczynski and V.~Ponomariev, {\it On the linearized
  einstein--cartan theory},  Ann. Inst. Henri Poincar{\'e} {\bf 21} (1974)
  89--95.

\bibitem{Iyer:1994ys}
V.~Iyer and R.~M. Wald, {\it {Some properties of Noether charge and a proposal
  for dynamical black hole entropy}},  Phys. Rev. {\bf D50} (1994) 846--864
  [\href{http://arXiv.org/abs/gr-qc/9403028}{{\tt gr-qc/9403028}}].

\bibitem{DePaoli:2018erh}
E.~De~Paoli and S.~Speziale, {\it {A gauge-invariant symplectic potential for
  tetrad general relativity}}, JHEP{\bf 07} (2018) 040  [\href{http://arXiv.org/abs/1804.09685}{{\tt
  1804.09685}}].

\bibitem{Chakraborty:2018qew}
S.~Chakraborty and R.~Dey, {\it {Noether Current, Black Hole Entropy and
  Spacetime Torsion}},  [\href{http://arXiv.org/abs/1806.05840}{{\tt
  1806.05840}}].

\bibitem{Tseytlin:1981nu}
A.~A. Tseytlin, {\it {On the Poincare and De Sitter Gauge Theories of Gravity
  With Propagating Torsion}},  Phys. Rev. {\bf D26} (1982) 3327.

\bibitem{Dislocation}
E.~Kr\"oner, {\it Description of dislocation distributions},  in {\em
  Dislocation Modelling of Physical Systems: Proceedings of the International
  Conference, Gainesville, Florida, USA, June 22-27, 1980} (M.~F. Ashby,
  R.~Bullough and C.~Hartley, eds.), Elsevier, 2017.

\bibitem{Hehl:1976kj}
F.~W. Hehl, P.~Von Der~Heyde, G.~D. Kerlick and J.~M. Nester, {\it {General
  Relativity with Spin and Torsion: Foundations and Prospects}},  Rev. Mod.
  Phys. {\bf 48} (1976) 393--416.

\bibitem{Hehl:2013qga} 
  F.~W.~Hehl, Y.~N.~Obukhov and D.~Puetzfeld,
  \emph{On Poincar\'e gauge theory of gravity, its equations of motion, and Gravity Probe B,}
  Phys.\ Lett.\ A {\bf 377}, 1775 (2013)  [1304.2769].
  
\bibitem{Bianchi:2013rya}
E.~Bianchi and A.~Satz, {\it {Mechanical laws of the Rindler horizon}},  Phys.
  Rev. {\bf D87} (2013), no.~12 124031
  [\href{http://arXiv.org/abs/1305.4986}{{\tt 1305.4986}}].

\bibitem{DeLorenzo:2017tgx}
T.~De~Lorenzo and A.~Perez, {\it {Light Cone Thermodynamics}},  Phys. Rev. {\bf
  D97} (2018), no.~4 044052 [\href{http://arXiv.org/abs/1707.00479}{{\tt
  1707.00479}}].

\end{thebibliography}
\end{document}